# Inexact Augmented Lagrangian Method-Based Full-waveform Inversion with Randomized Singular Value Decomposition

Jiahang Li[1], Hitoshi Mikada[1,2] and Junichi Takekawa[1]

[1]Department of Civil and Earth Resources Engineering, Kyoto University, C1-1-119

Kyotodaigaku-katsura, Nishikyo-ku, Kyoto, Japan

[2]Center for Integrated Research and Education of Natural Hazards, Shizuoka University,

Shizuoka Prefecture, Japan

**Corresponding Author: li.jiahang.n83@kyoto-u.jp**

**Keywords**: Full-waveform inversion

**SUMMARY**

Full Waveform Inversion (FWI) is a modeling algorithm used for seismic data processing and subsurface structure inversion. Theoretically, the main advantage of FWI is its ability to obtain useful subsurface structure information, such as velocity and density, from complex seismic data through inversion simulation. However, under complex conditions, FWI is difficult to achieve high-resolution imaging results, and most of the cases are due to random noise, initial model, or inversion parameters and so on. Therefore, we consider an effective image processing and dimension reduction tool, randomized singular value decomposition (rSVD) - weighted truncated nuclear norm regularization (WTNNR), for embedding FWI to achieve high-resolution imaging results. This algorithm obtains a truncated matrix approximating the original matrix by reducing the rank of the velocity increment matrix, thus achieving the truncation of noisy data, with the truncation range controlled by WTNNR. Subsequently, we employ an inexact augmented Lagrangian method (iALM) algorithm in the optimization to compress the solution space range, thus relaxing the dependence of FWI and rSVD-WTNNR on the initial model and accelerating the convergence rate of the objective function. We tested on two sets of synthetic data, and the results show that compared with traditional FWI, our method can more effectively suppress the impact of random noise, thus obtaining higher resolution and more accurate subsurface model information. Meanwhile, due to the introduction of iALM, our method also significantly improves the convergence rate. This work indicates that the combination of rSVD-WTNNR and FWI is an effective imaging strategy which can help to solve the challenges faced by traditional FWI.

# INTRODUCTION

Full Waveform Inversion (FWI) is a subsurface imaging technique in the field of seismic exploration, and this technique inverts and fits field or artificially synthesized seismic data to obtain the physical properties of underground media, such as velocity and density (Virieux & Operto 2009). The conventional workflow of FWI involves comparing the actual measured seismic wavefield with the computed seismic wavefield, calculating the difference between them through least squares, obtaining the updated direction based on the gradient method, and the velocity increment according to the updated direction, and superimposing the velocity increment with the result of the previous iteration to complete an iteration (Zhou *et al.* 2015). Ideally, the FWI algorithm will produce an inversion result very close to the real model after countless iterations, because, as a fitting algorithm, the inversion velocity model will gradually converge the real model under reasonable preconditions (Métivier *et al.* 2013). However, in reality, FWI still faces many challenges. For example, the errors inherent in the data collection equipment itself, or the impact of external environmental factors on the propagation of seismic waves may introduce random noise, which can significantly affect the effectiveness of FWI (Moghaddam & Herrmann 2010). In particular, high-level random noise can mask useful seismic information, making it impossible to capture some detailed underground information, and even slow down the convergence rate of FWI due to overfitting, resulting in erroneous inversion results (Aghamiry *et al.* 2021a). In addition, the unconstrained form of least squares, due to its overly loose solution space, often causes FWI convergence to easily fall into local minima. This problem is also often jointly considered with FWI's excessive dependence on the initial model, which has the same underlying mathematical logic, i.e., FWI, as a fitting algorithm, finds it difficult to accurately converge on non-strictly convex functions (Chi *et al.* 2014). Therefore, to achieve high-resolution and high-precision FWI, optimization of the inversion algorithm is crucial.

In the context of FWI, numerous approaches have been proposed to mitigate the potential overfitting issues encountered during the inversion process, such as Tikhonov regularization or Total Variation (TV) regularization (Song & Alkhalifah 2020). The overfitting issue frequently occurs due to the inherently noisy nature of real seismic data, often polluted with random noise or artifacts. After numerous iterations of FWI, especially with suboptimal initial models, the inversion process can be significantly degraded by this noise. In other words, the algorithm itself might mistakenly interpret the noise as genuine seismic signals. Incorporating them into the inversion process could lead to high sensitivity to noise, resulting in overfitting (Li *et al.* 2016). Tikhonov regularization tackles this by adding a regularization term to the objective function, striving to make the model as smooth as possible, which no longer merely minimizes the difference in the wavefield but also minimizes model complexity. This method seeks a balance between reducing error and preventing overfitting, suppressing the sensitivity of FWI to noise, and enhancing the algorithm's generalization capability (Aghamiry *et al.,* 2020). However, it is not ideal for dealing with highly discontinuous models or models with substantial velocity contrasts, because Tikhonov regularization generates smoother solutions. So, a clear manifestation of this is that it might blur the outline of salt bodies, resulting in reduced resolution (Zhu *et al.* 2021). Therefore, another regularization algorithm, TV regularization, has garnered attention owing to its introduction of sparse solutions. TV regularization allows the solution to have sharp boundaries and discontinuities, which is crucial for inverting various complex models in Earth Science (Qu *et al.* 2019). However, TV regularization itself also has notable issues. For instance, it involves non-smooth optimization problems (Aghamiry *et al.* 2021b).

Further, in FWI, the construction of sparse dictionaries based on the Singular Value

Decomposition (SVD) algorithm presents a novel approach to dealing with noise. SVD and sparse dictionary methods (Li & Harris 2018) can learn from a training dataset to build a highly adaptable dictionary capable of better handling complex, non-stationary, and non-fixed signals. Sparse encoding methods (Guo *et al.* 2020) can represent signals as sparse linear combinations, which can help isolate noise from signals, as signals can typically be represented by elements in the dictionary, while noise cannot (Liu *et al.* 2022). However, the conventional construction of sparse dictionaries and the solution of sparse encoding can require extensive computation, especially for large-scale seismic data, where both computational cost and memory consumption are crucial considerations. Additionally, the quality of the dictionary greatly affects the results. If the training data doesn't adequately represent the data to be processed, the effectiveness of the dictionary could be diminished (Wang *et al.* 2021).

Based on the question above, we propose to optimize the FWI iteration process with a combination of randomized singular value decomposition (rSVD)with weighted truncated nuclear norm regularization (WTNNR) and the inexact augmented Lagrangian method (iALM) algorithm to speed up the convergence rate of FWI. The most significant difference between the rSVD algorithm and conventional algorithms is that it first generates a compressed matrix relative to the original matrix through Gaussian random matrices and Economic QR decomposition. The scale of this compressed matrix is determined by the truncation coefficient $k$, which is embedded within the internal loop of FWI to achieve layered optimization of the velocity increment with each iteration, and the truncation parameter $k$ increases as the number of iterations increases. Using the rSVD algorithm can not only significantly reduce the scale of seismic data but also reduce memory consumption and computing time by adjusting the truncation coefficient (Liu & Peter 2020). More importantly, we embedded a WTNNR algorithm in rSVD to handle the Σ matrix (rectangular diagonal matrix with singular values) after rSVD decomposition. WTNNR is a special low-rank matrix recovery method that improves upon the traditional nuclear norm minimization method (Deng *et al.* 2020). In WTNNR, the singular values in the Σ matrix are sorted, and each singular value decomposed by rSVD is assigned a weight according to its size. Subsequently, according to a predetermined threshold, the singular values are truncated, i.e., singular values less than the threshold are set to 0, thereby acting as singular value shrinkage. The weighted singular values then replace the original singular values in the Σ matrix. The new Σ matrix and unitary matrixes are used for reconstruction to obtain the recovered low-rank matrix. The advantage of this method is that it can better balance the low-rank nature and data fidelity, as it can assign different weights to different singular values and can remove smaller singular values through the threshold, thus achieving a better denoising effect (Gu *et al.* 2017).

In this paper, we innovatively introduce the rSVD-WTNNR algorithm to improve the robustness of FWI for data sets with high background random noise. The acceleration in computation is realized by adopting the iALM algorithm. The synthetic model (2004 BP model) is tested by comparing the inversion result of the Tikhonov regularization FWI with those of rSVD-WTNNR FWI to verify the superiority of our method.

## RANDOMIZED SINGULAR VALUE DECOMPOSITION: RSVD

The full waveform inversion (FWI) as a fitting algorithm can be represented in the frequency domain as:

$$\min_{\chi} \frac{1}{2} \left\| Y - F[\chi, \mathcal{S}] \right\|_F^2, \tag{1}$$

where $Y$ is the input actual observed seismic data, $\chi$ is the model parameters, and $\mathcal{S}$ is the source matrix. The non-linear function $F$ is the seismic data obtained from the forward simulation of the equation in the frequency domain. In the Newton method, the function of equation 1 is minimized iteratively. Each of these iterations requires finding the inverse of the Hessian matrix, acting on the gradient, and then obtaining the velocity increment by choosing an appropriate iteration step:

$$\tilde{\chi} = H^\dagger g, \tag{2}$$

$$\chi = \chi + \alpha \, \tilde{\chi}, \tag{3}$$

where $H^\dagger$ is the approximation of the inverse of the Hessian matrix, $g$ is the gradient, $\alpha$ is the iteration step size, and $\chi$ is the model update. Equations 2 and 3 are classical algorithms for solving the inverse problem of the FWI using the Gauss-Newton method. However, the conventional algorithm is very computationally intensive because each of its iterations requires solving the Gauss-Newton subproblem, and each sub-iteration of the subproblem requires multiple solving of the wave equation (Li *et al.* 2012). At the same time, FWI updates the model by minimizing the misfit between observed and simulated data. At each iteration step, some corrections are made to the current model parameters. On the contrary, if the velocity increment is of poor quality, the model after each iteration may not be closer to the real model or even deviate from the real model, which will affect the accuracy of the FWI inversion results (Pan *et al.* 2016).

To achieve high-resolution FWI and reduce its computational burden, we optimize the model update by integrating dimensionality reduction techniques with image processing technologies. In FWI, we can reduce the rank of velocity increments through truncation, which might help to eliminate high-frequency, small-scale structural noise, as this type of noise often corresponds to smaller singular values. Therefore, by eliminating these smaller singular values, we can obtain a better and more stable velocity increment. Specifically, we first employ the rSVD algorithm to compress the increment distribution matrix in each iteration, and through the truncation parameter $k$. We obtain a smaller-scaled compressed matrix and then recover it to a similar matrix with the same size as the original one. However, different from the original matrix, this similar matrix has undergone truncation optimization. So, its rank is restricted by the truncation parameter $k$, making it significantly smaller than the rank of the original matrix. On the other hand, this truncated matrix only contains the most important features of the original matrix and discards smaller singular values, thus helping to remove noise and other unimportant features.

In this paper, we describe rSVD's corresponding steps concerning the FWI. We reshape the velocity increments $\tilde{\chi} \in \mathbb{R}^{m \times n}$ as primitive matrices, where $m$ and $n$ are the number of rows and columns, which we multiply by a Gaussian random matrix to compress them into a process matrix associated with a truncation parameter $k$:

$$\chi^{'} = \tilde{\chi}\,\Omega,\qquad(4)$$

where $\tilde{\chi} \in \mathbb{R}^{m \times k}$ is reshaped velocity increments and $\Omega$ is the Gaussian random matrix. After obtaining the process matrix $\chi'$, we subject it to an economic QR decomposition. The reason why we do not use regular QR decomposition is that the upper triangular matrix $R$ obtained by standard QR decomposition is redundant with size $R \in \mathbb{R}^{m \times k}$, which will be eventually discarded, while economic QR decomposition can quickly obtain the upper triangular matrix $R_e$ with size $R_e \in \mathbb{R}^{m \times k}$, which speeds up the overall operation:

$$Q = eqr(\chi'), \tag{5}$$

where $eqr$ is the economic QR decomposition (Song $et\ al.$ 2017). Next, we multiply $Q$ again with the velocity increments $\tilde{\chi}$ to get a preliminary compression matrix:

$$\chi_c = Q^* \tilde{\chi}, \tag{6}$$

where $Q^*$ is the transpose of matrices $Q$, and $\chi_c$ is the compression matrix we initially obtained. It is of size $\chi_c \in \mathbb{R}^{k \times n}$ compared to the original matrix of size $\tilde{\chi} \in \mathbb{R}^{m \times n}$. Here, the singular values of the approximated compression matrix are controlled by the stage parameters, i.e., no matter how large the increment of the original model is, we can find a compression matrix consisting of the singular values of the first $k$-rank and representing the main features of the original matrix (Halko $et\ al.$ 2011).

Subsequently, we need to perform a complete SVD of the small matrix $\chi_c$ to obtain the approximate singular values and singular vectors, obtained as follows:

$$\tilde{\psi} \Sigma_k V^* = \chi_c, \tag{7}$$

where $\tilde{\psi}$ is the left singular vector, $\Sigma_k$ contains the approximated first $k$ singular values in the diagonal elements, $V^*$ is the right singular vectors.

The above describes the decomposition process of velocity increment reshaping size in FWI by rSVD. Through these steps, we can quickly obtain the approximate singular values and singular vectors of the velocity increment. We provide a comparison of the computational speeds of rSVD and full SVD in Figure 1, which is based on the 2004 BP model. As shown in Figure 1, where the horizontal axis is the rank of the approximating matrix, and the vertical axis is the computational time, Figure 1(a) shows that, due to the smaller model size, the computational speed of rSVD is significantly superior to that of full SVD in the early stages of iteration, with a significant difference in their speed curves. Additionally, Figure 1(b), where the horizontal axis is the singular value order of the singular value matrix, and the vertical axis is the singular value, Figure 1(b) shows that with $k=20$ results in the original model, the curve of approximate singular values from rSVD also significantly differs from the singular value estimation curve from full SVD. The smaller model size corresponds to a smaller number and range of singular values, carrying out an initial truncation operation while retaining the main information of the model.

Subsequently, we recover the left singular vector $\tilde{\psi}$ and combine it with the singular value matrix $\tilde{\Sigma}_k$ that has been optimized by weighted truncated nuclear norm regularization (WTNNR) to obtain an output matrix of the same size as the input matrix $\tilde{\chi} \in \mathbb{R}^{m \times k}$, thus completing a single overall optimization:

$$\tilde{\chi} = \psi \, \tilde{\Sigma}_k V^*, \tag{8}$$

where $\tilde{\psi}$ is the recovered left singular vector, and $\tilde{\Sigma}_k$ is the singular value matrix optimized by WTNNR.

Moreover, the selection process for the truncation parameter is not difficult. We first set the initial value of $k$ before the forward modeling. We need to select the iteration step length $\varsigma$ with the interactions of the model increment so that the truncation parameter is linearly superimposed with the compensation $\varsigma$ in each inner loop. This procedure will result in a series of gradually expanded truncation ranges. Figure 2 shows the schematic diagram of the update process for the truncation parameter $k$. As $k$ increases, the singular value range of the approximate matrix gradually approaches that of the original matrix. Also, based on equations 1-8, we have provided a visualization flow chart of rSVD in FWI, as shown in Figure 3. Here, we use the velocity model of the 2004 BP model of size $124 \times 124$ as an example of velocity increment, providing a visual reference for one time's rSVD optimization process, in which the truncation parameter is taken as $k=50$ as an instance.

**WEIGHTED TRUNCATED NUCLEAR NORM REGULARIZATION: WTNNR**

FWI uses gradients to update the velocity model, and the conventional second-order Gauss-Newton method uses the inverse of the Hessian matrix to scale the gradient. However, these types of algorithms have a large computational load (Horst *et al.* 2000). By adopting dimensionality reduction techniques, the computational load of FWI can be significantly compressed, especially as an iterative algorithm. In this study, we leverage the truncation property of rSVD and WTNNR to optimize the iterative update items for enhancing sparsity and thereby reducing computational load:

$$\delta\chi = \arg\min_{\delta\chi} \left\| \tilde{\chi} \right\|_{WT}, \qquad (9)$$

where $\left\| \bullet \right\|_{WT}$ is one kind of truncated nuclear norm, and $\delta\chi$ is the optimized velocity increment. We can transform the WTNNR problem into the following form:

$$\left\| \tilde{\chi} \right\|_{WT} = \left\| \tilde{\chi} \right\|_* - \max_{\psi\psi^k = I, VV^k = I} Trace(\psi_k \tilde{\chi}(\tilde{\Sigma_k})V_k^T) \qquad (10)$$

where $\psi_k \in \mathbb{R}^{k \times k}$ and $V_k \in \mathbb{R}^{k \times n}$ satisfying $\psi_k \psi_k^T = I_{k \times k}$ and $V_k V_k^T = I_{k \times k}$, the truncation parameter $k$ always satisfies $k \leq \min(m, n)$, $Trace(\bullet)$ represents the trace of the matrix. For the singular value matrix obtained after rSVD in the previous section, we have to perform windowing. After the windowing process, more non-zero singular values are removed, which dramatically reduces the size and complexity of the matrix for further truncation operations:

$$W(\tilde{\Sigma_k}) = \max(\tilde{\Sigma_k} - \frac{W_k}{2}, 0), \qquad (11)$$

where $W_k$ is the truncation matrix, $W(\tilde{\Sigma_k})$ is the new singular value matrix after the weighting process. If $\frac{W_k}{2} \geq \tilde{\Sigma_k}$, $W(\tilde{\Sigma_k})$ is taken to be 0. We give the schematic in Figure 4. The blue solid line is the value of the diagonal elements of the singular value matrix of the full SVD. The red solid line is the value of the singular value matrix after the first step optimization with rSVD processing. In contrast, the black solid line is the value of the singular value matrix of the weight we constructed, and, finally, the violet solid line is the value of the diagonal elements of the new singular value matrix filtered with the weight window values. Compared to the original matrix, the values of the noisy singular value matrix have been optimized with two times truncation and singular value shrinkage optimization. In addition, we give a visual explanation in Figure 5. Figures 5(a-c) are the matrixes with a 9:9 horizontal-to-vertical ratio. Figure 5(a) is the singular value matrix after rSVD, Figure 5(b) is the weighted matrix we constructed, and Figure 5(c) is the singular value shrinkage matrix obtained after the operation according to equation 11. In addition, Figures 5(d-f) are the same as described above, except that the ratio of matrix sizes is $19 \times 19$. We can intuitively find that all of the singular value matrices after singular value shrinkage treatment have smaller magnitudes and fewer non-zero elements and implemented singular value shrinkage optimization.

The choice of the weight vector is crucial. According to a priori knowledge, larger singular values are more important than smaller ones, especially in denoising applications. The larger the singular value, the less it needs to be shrunk (Chang *et al.* 2000), so the idea is that the weight assigned to the *i*-th singular value is inversely proportional to it:

$$W_i = \beta \sqrt{\gamma} / (\sigma_i + \varepsilon), \tag{12}$$

where $\beta$ is a nonnegative constant, $\sigma_i$ is the locally estimated variance at the *i*-th position, $\gamma$ is the number of similar patches, and $\varepsilon = 2^{-52}$ to avoid the divisor being zero:

$$\hat{\sigma_i} = \sqrt{\max(\sigma_i^2 - \gamma \sigma_k^2, 0)}, \tag{13}$$

where $\sigma_i$ is the *i*-the singular value of the matrix. Similarly, in the case of $\gamma \sigma_k^2 \geq \sigma_i^2$, $\hat{\sigma_i}$ is taken to be 0. The four sets of singular value matrices shown in Figure 6 are derived from the original velocity increments, the singular value matrix of rSVD, the weight shrinkage WTNNR, and the optimized singular value matrix. We initially tested the effect of the treatment with and without WTNNR in Figure 7, and it can be seen from the results of a single rSVD treatment. Especially in the case of $k = 1$, the result with-WTNNR is better optimized, which highlights the main features of the intermediate salt-column portion and shrinks the non-salt-column portion on both sides.

**INEXACT AUGMENTED LAGRANGIAN METHOD: IALM**

Further, FWI can be expressed as the following regularized PDE-constrained optimization form in the frequency domain (van Leeuwen & Herrmann 2013) and the constraint form based on augmented Lagrangian:

$$\min_{\chi,u} \max_{\lambda} L(\chi,\tau) = \min_{\chi,u} \max_{\lambda} \left\| Pu - Y \right\|_2^2 + (\hat{\lambda}_j)^T \left[ A(\chi)u - b \right] + \tau \left\| A(\chi)u - b \right\|_2^2, \tag{14}$$

where $\left\| \bullet \right\|_2^2$ is the Euclidean norm, $u \in \mathbb{R}^{Na \times 1}$ is the model wavefield, $Y \in \mathbb{R}^{Ma \times 1}$ is the recorded seismic data, $b \in \mathbb{R}^{Na \times 1}$ is the source term, and the linear observation operator $P \in \mathbb{R}^{Ma \times Na}$ sampling $u$ at the receiver positions, $\chi \in \mathbb{R}^{Na \times 1}$ is the model parameters, which contains preliminary information about underground parameters. The $A(\chi) \in \mathbb{R}^{Na \times Na}$ is the discretized PDE, which is synergic with the $\chi$. Moreover, the $Na = N_x \times N_z \times N_t$, $Ma = N_t \times N_r$, where $N_x$ and $N_z$ represent the number of grid points for sampling the numerical model in the horizontal and vertical directions, respectively, $N_r$ is the number of receivers, $N_t$ is the number of time samples (Gholami *et al.* 2022). Equation 14 can be accelerated through some effective acceleration techniques initialized by Nesterov (Sahin *et al.* 2019):

$$\lambda_j = \hat{\lambda}_j - \tau(A(\chi)u - b), \tag{15}$$

$$t_{j+1} = (1 + \sqrt{1 + 4(t_j)^2}) / 2, \tag{16}$$

$$\hat{\lambda}_{j+1} = \lambda_j + \frac{t_j - 1}{t_j + 1}(\lambda_j - \lambda_{j-1}) + \frac{t_j}{t_{j+1}}(\lambda_j - \hat{\lambda}_j), \tag{17}$$

where $\hat{\lambda}_1 = \lambda_0$, and $t_1 = 1$. iALM is a new iterative method for solving the linear constraint convex function minimization problem of equation 14. This iterative method is an inexact version of the augmented Lagrange method (ALM). Since every single sub-problem needs to be solved accurately in each external iteration, the computational cost of the other variants version of ALM still has the potential to be reduced. More importantly, the sub-problems of accelerated augmentation Lagrangian do not have closed-form solutions; therefore, the algorithm we adopted allows inexact solutions to the sub-problems. This point makes the iALM accelerate the convergence rate and reduce the computational cost. It overcomes the problem that the exact solution to the sub-problem requires much calculation. Equation17 can also be simplified to:

$$\hat{\lambda}_{j+1} = \lambda_j + \frac{t_j - 1}{t_j + 1}(\lambda_j - \lambda_{j-1}). \tag{18}$$

Finally, as shown in Algorithm 1, we give the computational flow of FWI based on the rSVD-WTNNR and iALM. As shown in Figure 8, we provide the full updated workflow for improving FWI.

**SYNTHETIC EXAMPLES**

To test the performance of the improved FWI proposed in this paper, we compared the results of the Tikhonov regularization-based FWI and the improved FWI using a synthetic data set under three different random noise disturbances. For the forward modeling part, we adopted a frequency-domain multiscale algorithm (Bunks *et al.* 1995), combined with the Perfectly

Matched Layer (PML) as the boundary condition (Komatitsch & Tromp 2003). Moreover, when comparing the two algorithms, we used model error to quantify the inversion results for comparison:

$$\left\| \mathcal{M}_{true} - \mathcal{M}_{inv} \right\|_2 / \left\| \mathcal{M}_{true} \right\|_2 . \tag{19}$$

where the $\mathcal{M}_{true}$ and $\mathcal{M}_{inv}$ represent the actual model and inversion result, respectively (Warner & Guasch 2016).

The main objective of this paper is to propose a robust FWI algorithm; therefore, we add random noise to the synthetic data, thus increasing the inversion difficulty of FWI. Then, we test the ability of improved FWI to achieve high-resolution imaging in the presence of random noise interference. We use the signal-to-noise ratio formula to define the level of random noise:

$$\mathrm{SNR} = 20 * \mathrm{Log}_{10} (\frac{\left\| \mathcal{D} \right\|_2}{\left\| \eta \right\|_2}), \tag{20}$$

where the $\eta$ is the noise data, the $\mathcal{D}$ is the signal data.

**2004 BP MODEL**

The famous 2004 BP model is divided into three parts, and we adopted the middle part of the model, which consists of a high-speed salt body and channels on both sides. This model simulates the geological features of the eastern/central Gulf of Mexico and offshore Angola, which is a very important geological model in applied geophysical research. The challenge of inverting this model lies in outlining the contours of the central salt body, capturing the deep salt pillar, and restoring the channels. Due to the large velocity difference between the salt body and the surrounding model, high-precision inversion of this model is also one of the challenges of FWI (Billette & Brandsberg-Dahl 2005).

We present the true velocity model in Figure 9 (a) and the initial velocity model in Figure 9 (b) of the 2004 BP model in Figure 9. In the synthetic data, we tested the inversion results under three different noise levels. So, we first give the data sets for three different $dB$s of random noise, as shown in Figure 10. Figure 10 (a) is the noise-free data. Figures 10 (b-d) are the data sets for three different random noises of 8 $dB$,12 $dB$, and 16 $dB$, respectively. Figures 10 (e-g) are the data sets after adding the random background noise sets. Subsequently, we present the velocity increment with many different truncation parameters under 12 $dB$ noise, as shown in Figure 11, with the truncation parameters being $k = 10$, $k = 20$, $k = 30$, $k = 40$, and $k = 50$, respectively. Figures 11 (a1-a4) show the conventional velocity increment models near the frequency of 4.5 Hz. Figures 11 (b1-b4) show the modified velocity increment models near the frequency of 4.5 Hz. As can be seen, as the truncation parameter increases from $k = 10$ to $k = 50$, due to the gradual increase in the rank of the singular value matrix, the order of the eigenvectors of the velocity increments also increases gradually. The energy distribution of the eigenvectors gradually moves from the top to the bottom of the subsurface model. The eigenvectors gradually shift from representing the main outline of the subsurface structure to representing the overall structure. Figure 12 presents the inversion results and comparisons under 8 $dB$ noise conditions. Specifically, Figures 12 (a1-a5) show the inversion results of the FWI based on Tikhonov regularization at frequencies of 2.15 Hz, 3.9 Hz, 5.35 Hz, 7.70 Hz, and 9.24 Hz, respectively. Figures 12 (b1-b5) show the differences between the inversion models based on Tikhonov regularization FWI and the true velocity models. Figures 12 (c1-c5) present the inversion results of the improved FWI proposed in this paper. Figures 12 (d1-d5) show the differences between the inversion models based on the enhanced FWI proposed in this paper and the true velocity model. To compare the inversion results of the two algorithms more clearly, we present the one-dimensional velocity model comparisons in both horizontal and vertical directions in Figure 13. Specifically, Figures 13 (a-f) show the vertical velocity comparisons at six different *x*-positions, while Figures 13 (g-l) present the horizontal velocity comparisons at six different *y*-positions. We have thoroughly compared the inversion results of the two algorithms. As can be seen from Figure 13, the results of the modified FWI proposed in this paper are closer to the true velocity model than those of the Tikhonov FWI. Subsequently, in Figure 14, we present the inversion results under 12 *dB* random noise, and in Figure 15, we provide the comparisons of the one-dimensional velocity models of the two algorithms. Similar to the results under 8 *dB*, the inversion results of the improved algorithm proposed in this paper also outperform those of the Tikhonov FWI, which is manifested in better suppression of outliers, clearer delineation of the model contours, and smoother fitting effects for continuous media. Then, we present the inversion results under 16 *dB* noise conditions. Figure 16 shows the inversion results at five different inversion frequencies and their differences from the true velocity model. The comparisons of the one-dimensional velocity models are given in Figure 17, which includes six groups of vertical velocity comparisons and six groups of horizontal velocity comparisons. From the inversion results and comparison results, it can be seen that the inversion model based on the modified FWI proposed in this paper is closer to the true velocity model, and the velocity curves are also more similar. Finally, in Figure 18, we present the comparison curves of the misfit error and model error curves under three different noise conditions. The results show that regardless of the noise conditions, the convergence rate of the improved algorithm is faster, and the model error of the enhanced algorithm is also smaller compared to the Tikhonov-based FWI.

The inversion results under the three different noise conditions all prove that our proposed rSVD-WTNNR-based FWI has higher precision inversion results, better noise suppression capabilities, and faster convergence speed, which validate the effectiveness and feasibility of the improved algorithm proposed in this paper.

## DISCUSSIONS

FWI is a time-tested geophysical fitting algorithm. While its core framework has been widely accepted, there is still potential for improvement and refinement in its specific details, particularly when dealing with challenges found in practical applications. In pursuing a high-resolution, noise-resistant FWI, we have attempted to incorporate specific image processing techniques into the FWI framework. By combining well-established image processing algorithms with the overall framework of FWI, we aim to achieve higher resolution and better noise resistance.

Specifically, our approach originates from the treatment of velocity increment in FWI. By resizing the reconstruction vector, we convert the original problem of FWI velocity increment processing into an image denoising issue. For example, as illustrated in Figure 19, we start with a velocity increment in Figure 19 (a), which undergoes SVD to yield Figure 19 (b). We then add a random noise in Figure 19 (c) to Figure 19 (a), resulting in a new singular value matrix in Figure 19 (d). How we handle the singular value matrix Figure 19 (d) is similar to how we take the objective function in FWI; hence, we consider this optimization method a regularization strategy. Back-calculating Figure 19 (d) gives us the velocity increment, as seen in Figure 19 (e). If we simply truncate Figure 19 (d), we remove all singular values within the red box and back-calculate to end with Figure 19 (f). Since the result of Figure 19 (f) is significantly superior to those of Figure 19 (e), we believe this innovative approach provides us with new tools, strategies, and perspectives in dealing with FWI problems. Additionally, our proposed method potentially holds substantial value in addressing the ill-posed nature of FWI, especially when handling data noise and uncertainty in velocity increment.

Based on this, we consider employing a rSVD algorithm for preliminary truncation processing. rSVD is a potent tool for matrix decomposition, applicable to any real or complex matrix, and often used to address matrix decomposition problems in large-scale data. To elucidate why we adopt rSVD more intuitively, we interpret SVD from a geometrical perspective, as illustrated in Figure 20. SVD decomposes a target matrix into two orthogonal matrices and one diagonal matrix. These two orthogonal matrices represent two rotations, while the diagonal matrix $\Sigma$ embodies the degree of stretching or compression of the singular value matrix along the coordinate axes. The singular value matrix can also be interpreted as the 'importance' or 'degree of variation' of the parameters in the corresponding dimensions.

Furthermore, as shown in Figure 21, if we choose to retain only the top $k$ largest singular values, this is equivalent to preserving the information of the parameters in the most crucial $k$ dimensions and discarding the information in other dimensions. Hence, we have dimensionally reduced the velocity increment from the original $n$ to $k$ dimensions. Geometrically, this operation can be considered as projection, where we project the data onto a subspace constituted by the most significant $k$ dimensions. Additionally, in the practical processing of FWI, there often exists some noise. This noise generally distributes across all dimensions, but its influence is relatively minor in the main dimensions. So, when we perform dimensionality reduction via SVD, we can discard some smaller singular values, thus eliminating information in some minor dimensions, which often contain most of the noise. Therefore, SVD dimensionality reduction can effectively remove noise, enhancing the quality of inversion. However, for large-scale matrices, traditional SVD algorithms might become computationally very expensive. To tackle this problem, we propose employing the rSVD algorithm in FWI. rSVD is a singular value decomposition algorithm based on randomization technology, capable of effectively dealing with large-scale data. Compared to traditional SVD, rSVD is more computationally efficient.

In this study, we innovatively introduce the combination of rSVD and weighted truncated nuclear norm regularization (WTNNR); the successful applications in other fields give us confidence in its potential in seismic exploration. However, it is worth noting that the optimality and stability of this method still need further research, especially the selection of truncation parameter $k$ and truncation step size $\varsigma$ are crucial. In addition, we have proposed a new method for improving the least squares objective function. We employed the inexact augmented Lagrangian method (iALM) algorithm to accelerate the convergence rate of the objective function. Although a large amount of mathematical literature has demonstrated its superior convergence characteristics in theory and achieved good results in experiments, there is still no definitive understanding of its parameter selection strategy and its impact on inversion results. Even though parameter selection methods, including Heuristic Search, Cross-Validation, and even Meta-Learning, have been proposed (Andrychowicz *et al.* 2016), there is still no perfect solution to this type of parameter selection problem; hence, this is also one of the issues that deserves our in-depth research in the future.

**CONCLUSIONS**

In response to the traditional FWI challenges of low resolution and outlier sensitivity, this paper introduces a novel optimization workflow to enhance FWI's inversion precision and robustness to noise under equivalent conditions. Within the FWI framework, we innovatively incorporate the rSVD algorithm into the inversion process and employ the WTNNR algorithm to optimize the singular value matrix of velocity increments. Subsequently, we utilize the iALM algorithm, which introduces variable Lagrange multipliers, thereby enhancing the stability and reliability of the FWI and accelerating its convergence rate. In the numerical experiment section, we conducted a comparative analysis of the inversion results derived from the improved FWI and traditional FWI for the 2004 BP model under three distinct low signal-to-noise ratio scenarios. The experimental outcomes demonstrate that the methodology proposed in this paper effectively mitigates the interference of outliers on inversion by truncating the eigenvalue range of velocity increments under various low signal-to-noise ratio conditions. This enhancement facilitates a more robust representation of model feature information and enables high-resolution imaging of complex structural deep regions.

**ACKNOWLEDGEMENTS**

**Figure 1.** (a) The blue solid line is the computation time for the full SVD and the red solid line is the processing time for the rSVD; (b) the blue solid line is the full SVD estimated singular value curve and the red solid line is the approximate singular value curve based on rSVD.

**Figure 2.** Schematic representation of the increase in the truncation parameter with the number of internal iterations, where $k$ is the truncation parameter and $\varsigma$ is the step size.

**Figure 3.** Visualization flowchart for rSVD, where $k = 50$ is the truncation parameter and the model size is 124 × 124.

**Figure 4.** The blue solid line is the full SVD estimated singular value curve, the red solid line is the approximate singular value curve based on rSVD, the black solid line is the weight curve based on WTNNR, and the purple solid line is the approximate singular value curve optimized by rSVD and WTNNR.

**Figure 5.** (a-c) 9 × 9 test data, (a) singular value matrix after rSVD decomposition; (b) weighted singular value matrix constructed based on WTNNR with singular values growing in the opposite direction of (a); and (c) optimized singular value matrix obtained according to equation 11. (d-f) 19 × 19 test matrices, (d) Singular value matrix after rSVD decomposition; (e) weighted singular value matrix constructed based on WTNNR with singular values growing in the opposite direction to (d); (f) optimized singular value matrix obtained according to equation 11.

**Figure 6.** 124 × 124 singular value matrix of the reconstructed velocity increment, (a) input matrix; (b) singular value matrix after rSVD-based optimization; (c) inverse weight matrix constructed based on WTNNR; and (d) singular value matrix of the velocity increment after rSVD-WTNNR optimization.

**Figure 7.** (a-c) Test results of velocity increment based on rSVD optimization without-WTNNR with truncation parameters *k=1, k=5,* and *k=10*; (d-f) test results of velocity increment based on rSVD optimization with-WTNNR. WTNNR can effectively enhance the features of the salt columns and slow down the disturbance of the channels on both sides.

**Algorithm 1.** FWI based on the rSVD-WTNNR & iALM.

**Figure 8.** Flowchart of accelerated augmented Lagrangian full-waveform inversion based on truncated randomized singular value decomposition in the frequency domain.

**Figure 9.** 2004 BP model, (a) true velocity model; (b) initial velocity model.

**Figure 10.** The real part of the 3 Hz data for the 2004 BP model. (a) noise-free data; (b-d) random noise of 8 *dB*, 12 *dB*, and 16 *dB*; (e-g) noisy data.

**Figure 11.** Conventional velocity increment model, (a1-a5) close to the 4.5 Hz frequency, the truncation parameter *k*, which increases as the number of internal iterations increases, is equal to 10, 20, 30, 40, and 50, respectively; modified velocity increment model, (b1-b5) close to the 4.5 Hz frequency, the truncation parameter *k* is equal to 10, 20, 30, 40, and 50, respectively.

**Figure 12.** 2004 BP model with 8 *dB* random background noise, (a1-a5) inversion results based on Tikhonov regularised FWI with frequencies of 2.15 Hz, 3.9 Hz, 5.35 Hz, 7.70 Hz, and 9.24 Hz, respectively; (b1-b5) velocity differences between the inversion results based on Tikhonov regularised FWI and the true velocity model; (c1-c5) inversion results based on modified FWI with frequencies of 2.15 Hz, 3.9 Hz, 5.35 Hz, 7.70 Hz, and 9.24 Hz, respectively; (d1-d5) velocity differences between the inversion results based on modified FWI and the true velocity model.

**Figure 13.** 2004 BP model with 8 *dB* random background noise, (a-f) the vertical comparison of one-dimensional velocity models at different *x*-positions, (a) $x$ = 2.84 km, (b) $x$ = 5.75 km, (c) $x$ = 11.08 km, (d) $x$ = 13.92 km, (e) $x$ = 15.00 km, (f) $x$ = 17.80 km; (g-l) the horizontal comparison of one-dimensional velocity models at different *y*-positions, (g) $y$ = 2.36 km, (h) $y$ = 2.76 km, (i) $y$ = 3.36 km, (j) $y$ = 3.68 km, (k) $y$ = 4.08 km, (l) $y$ = 4.56 km; where the actual velocity model is the solid black line, initial velocity model is grey dotted line, the Tikhonov FWI is solid blue line, and the modified FWI is solid red line.

**Figure 14.** 2004 BP model with 12 *dB* random background noise, (a1-a5) inversion results based on Tikhonov regularised FWI with frequencies of 2.15 Hz, 3.9 Hz, 5.35 Hz, 7.70 Hz, and 9.24 Hz, respectively; (b1-b5) velocity differences between the inversion results based on Tikhonov regularised FWI and the true velocity model; (c1-c5) inversion results based on modified FWI with frequencies of 2.15 Hz, 3.9 Hz, 5.35 Hz, 7.70 Hz, and 9.24 Hz, respectively; (d1-d5) velocity differences between the inversion results based on modified FWI and the true velocity model.

**Figure 15.** 2004 BP model with 12 *dB* random background noise, (a-f) the vertical comparison of one-dimensional velocity models at different *x*-positions, (a) $x$ = 4.48 km, (b) $x$ = 6.00 km, (c) $x$ = 11.08 km, (d) $x$ = 13.88 km, (e) $x$ = 14.68 km, (f) $x$ = 17.76 km; (g-l) the horizontal comparison of one-dimensional velocity models at different *y*-positions, (g) $y$ = 2.80 km, (h) $y$ = 3.36 km, (i) $y$ = 3.80 km, (j) $y$ = 4.76 km, (k) $y$ = 4.84 km, (l) $y$ = 5.28 km; where the actual velocity model is the solid black line, initial velocity model is grey dotted line, the Tikhonov FWI is solid blue line, and the modified FWI is solid red line.

**Figure 16.** 2004 BP model with 16 *dB* random background noise, (a1-a5) inversion results based on Tikhonov regularised FWI with frequencies of 2.15 Hz, 3.9 Hz, 5.35 Hz, 7.70 Hz, and 9.24 Hz, respectively; (b1-b5) velocity differences between the inversion results based on Tikhonov

regularised FWI and the true velocity model; (c1-c5) inversion results based on modified FWI with frequencies of 2.15 Hz, 3.9 Hz, 5.35 Hz, 7.70 Hz, and 9.24 Hz, respectively; (d1-d5) velocity differences between the inversion results based on modified FWI and the true velocity model.

**Figure 17.** 2004 BP model with 16 *dB* random background noise, (a-f) the vertical comparison of one-dimensional velocity models at different *x*-positions, (a) $x = 6.44$ km, (b) $x = 9.24$ km, (c) $x = 10.44$ km, (d) $x = 12.12$ km, (e) $x = 16.20$ km, (f) $x = 18.00$ km; (g-l) the horizontal comparison of one-dimensional velocity models at different *y*-positions, (g) $y = 2.68$ km, (h) $y = 2.88$ km, (i) $y = 3.84$ km, (j) $y = 4.08$ km, (k) $y = 4.80$ km, (l) $y = 5.44$ km; where the actual velocity model is the solid black line, initial velocity model is grey dotted line, the Tikhonov FWI is solid blue line, and the modified FWI is solid red line.

**Figure 18.** Model curves for the test, (a-b) misfit error and model error in the case of 8 *dB* random background noise; (c-d) misfit error and model error in the case of 12 *dB* random background noise; (e-f) misfit error and model error in the case of 16 *dB* random background noise; the solid red line represents the modified FWI, and the blue line represents the Tikhonov FWI.

**Figure 19.** (a) The initial test velocity increment; (b) the singular value matrix for this velocity increment; (c) 10 *dB* of random noise; (d) the test velocity increment after adding the noise disturbance; (e) the velocity increment after reconstructing the singular value matrix (d); (f) the new singular value matrix after simply truncating the singular value matrix (d); (g) the new velocity increment with a reconstructing of the singular value matrix (f); and (h) the difference between velocity increments (e) and (g).

**Figure 20.** Geometric interpretation of SVD, (a) target matrix; (b) left orthogonal matrix representing rotation; (c) singular value matrix representing longitudinal and horizontal stretching; (d) right orthogonal matrix representing rotation.

**Figure 21.** Geometric interpretation of the truncation operation of the SVD, (a) the target matrix; (b) the left orthogonal matrix represents the rotation; (c) the singular value matrix represents the longitudinal and horizontal stretching and truncation of its smallest singular value results in a longitudinal stretch of 0; (d) the right orthogonal matrix represents the rotation.

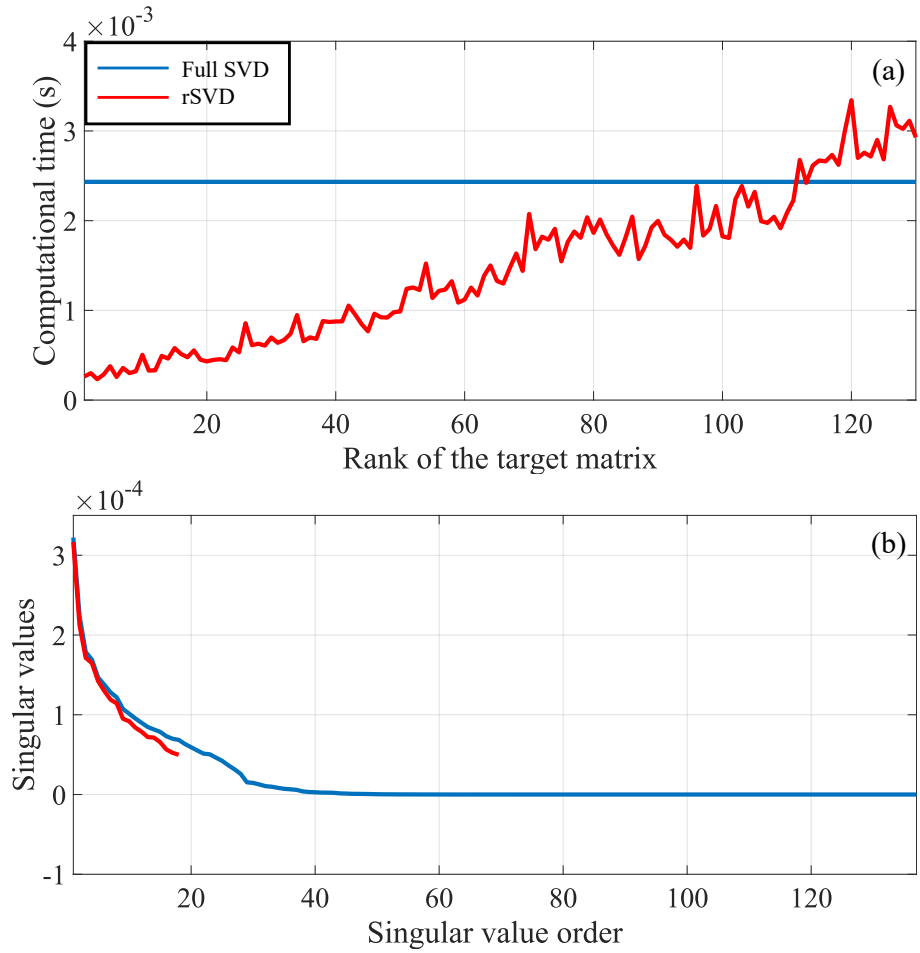

**Figure 1.** (a) The blue solid line is the computation time for the full SVD and the red solid line is the processing time for the rSVD; (b) the blue solid line is the full SVD estimated singular value curve and the red solid line is the approximate singular value curve based on rSVD.

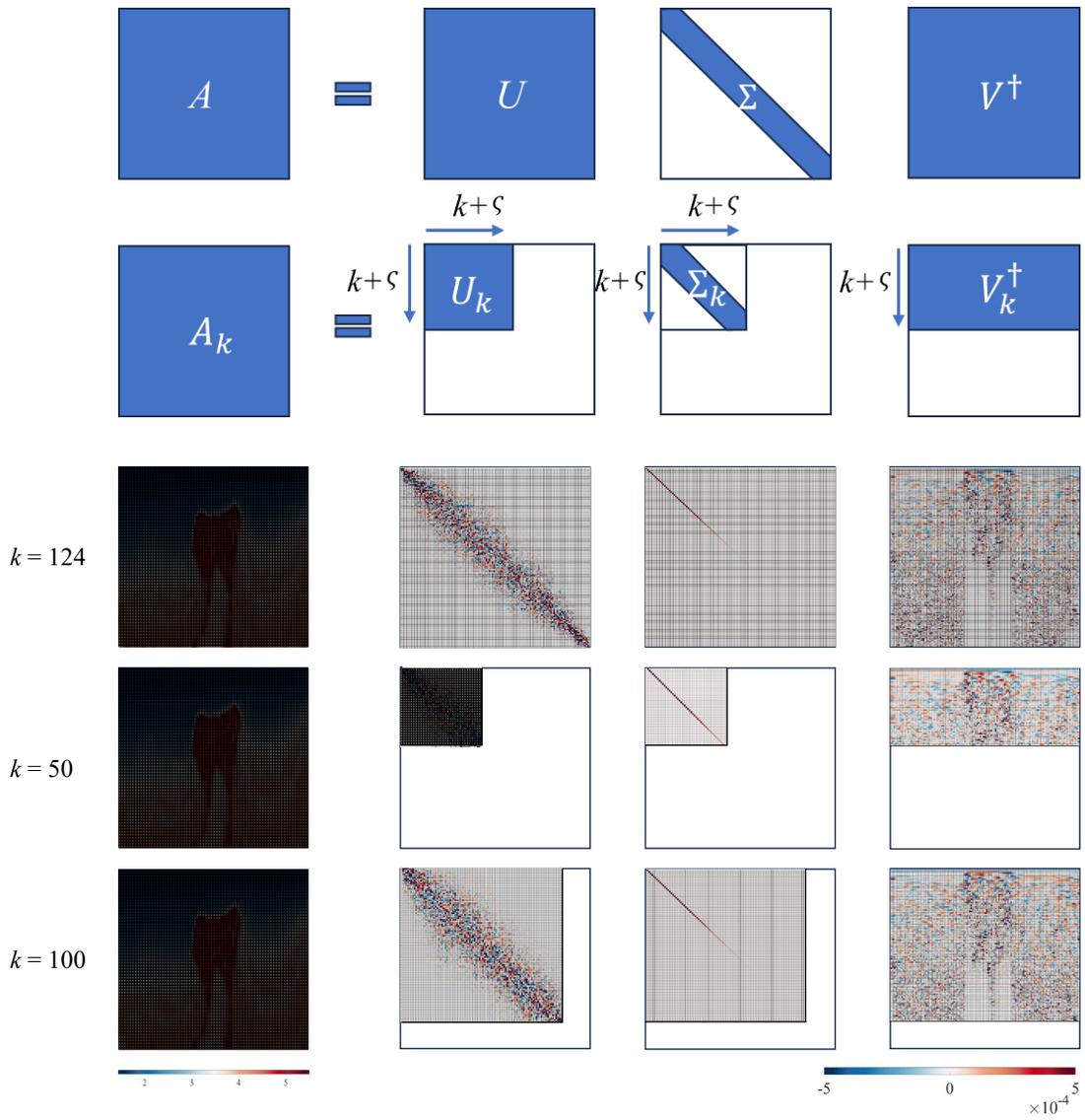

**Figure 2.** Schematic representation of the increase in the truncation parameter with the number of internal iterations, where $k$ is the truncation parameter and $\varsigma$ is the step size.

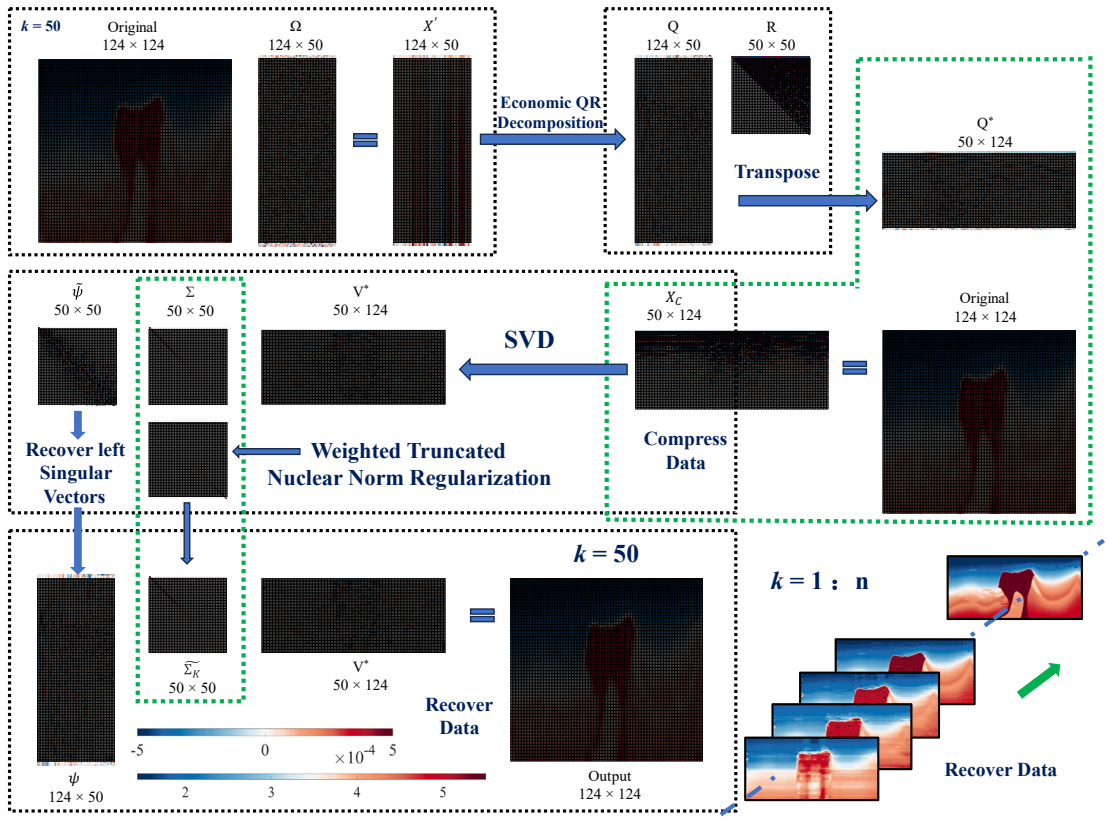

**Figure 3.** Visualization flowchart for rSVD, where *k = 50* is the truncation parameter and the model size is 124 × 124.

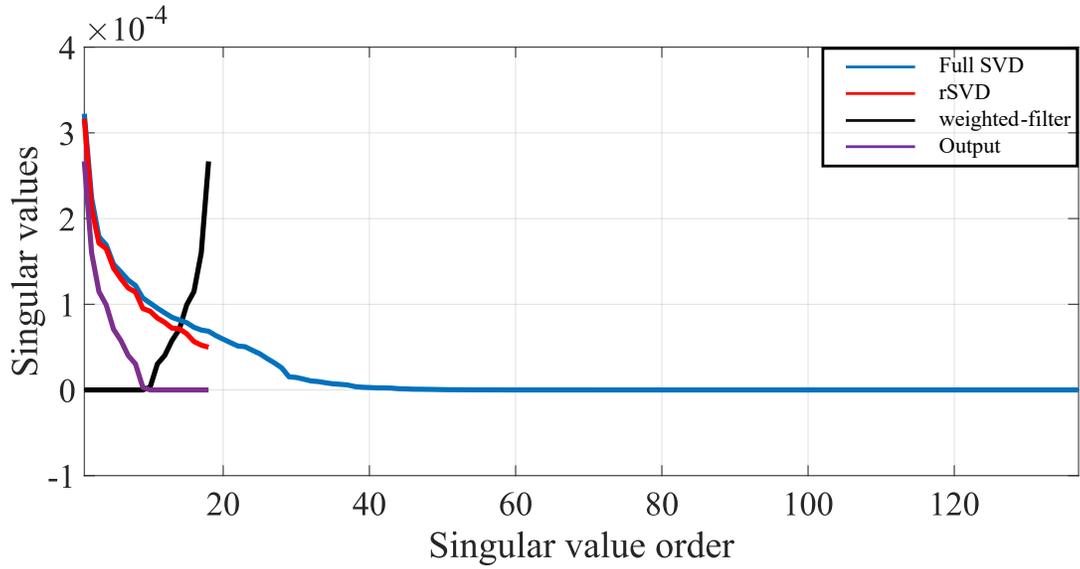

**Figure 4.** The blue solid line is the full SVD estimated singular value curve, the red solid line is the approximate singular value curve based on rSVD, the black solid line is the weight curve based on WTNNR, and the purple solid line is the approximate singular value curve optimized by rSVD and WTNNR.

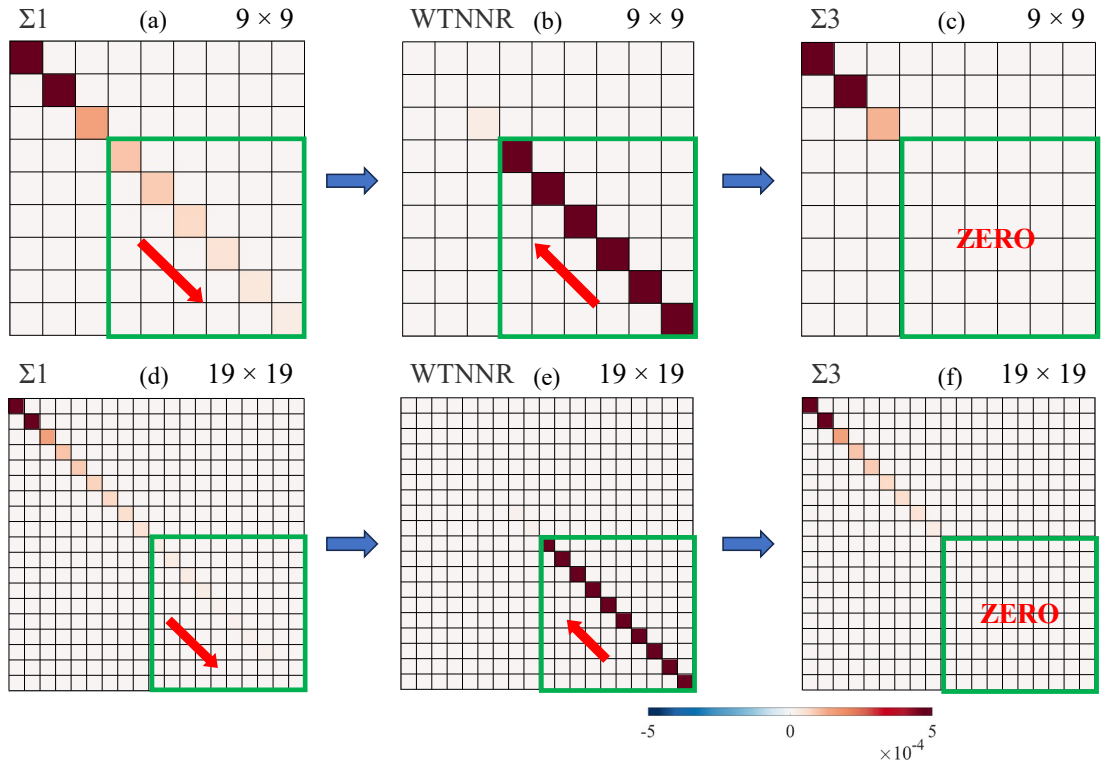

**Figure 5.** (a-c) 9 × 9 test data, (a) singular value matrix after rSVD decomposition; (b) weighted singular value matrix constructed based on WTNNR with singular values growing in the opposite direction of (a); and (c) optimized singular value matrix obtained according to equation 11. (d-f) 19 × 19 test matrices, (d) Singular value matrix after rSVD decomposition; (e) weighted singular value matrix constructed based on WTNNR with singular values growing in the opposite direction to (d); (f) optimized singular value matrix obtained according to equation 11.

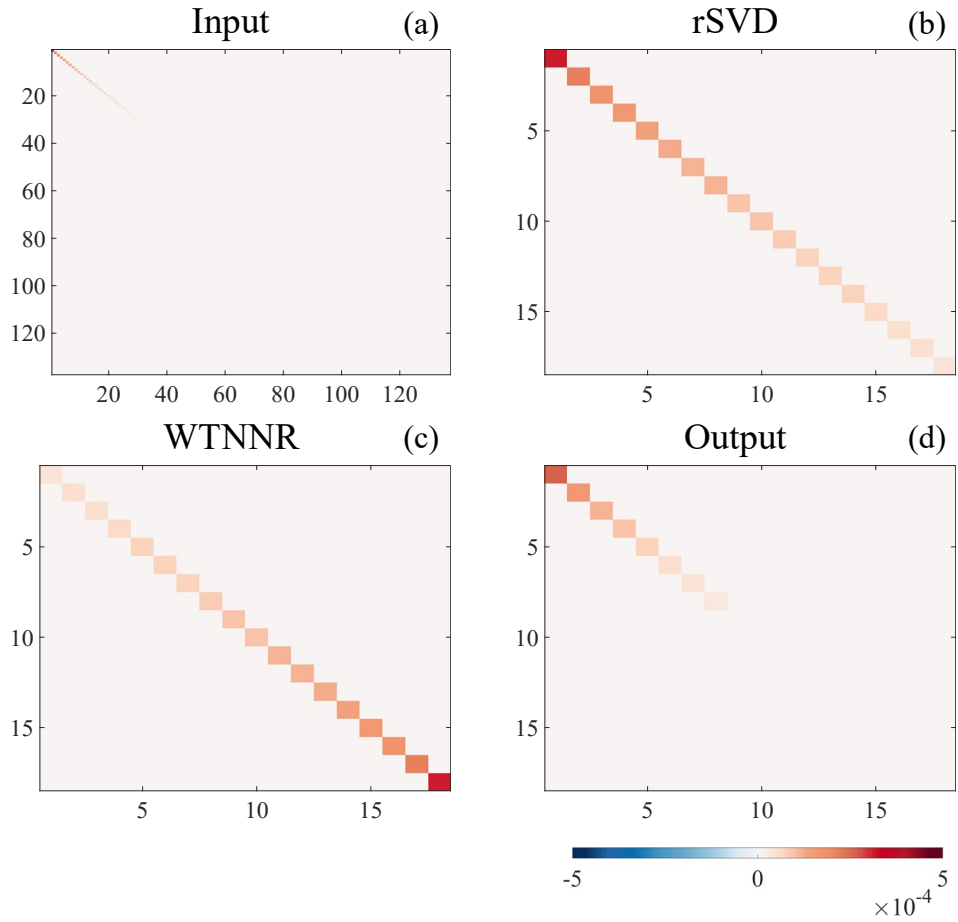

**Figure 6.** 124 × 124 singular value matrix of the reconstructed velocity increment, (a) input matrix; (b) singular value matrix after rSVD-based optimization; (c) inverse weight matrix constructed based on WTNNR; and (d) singular value matrix of the velocity increment after rSVD-WTNNR optimization.

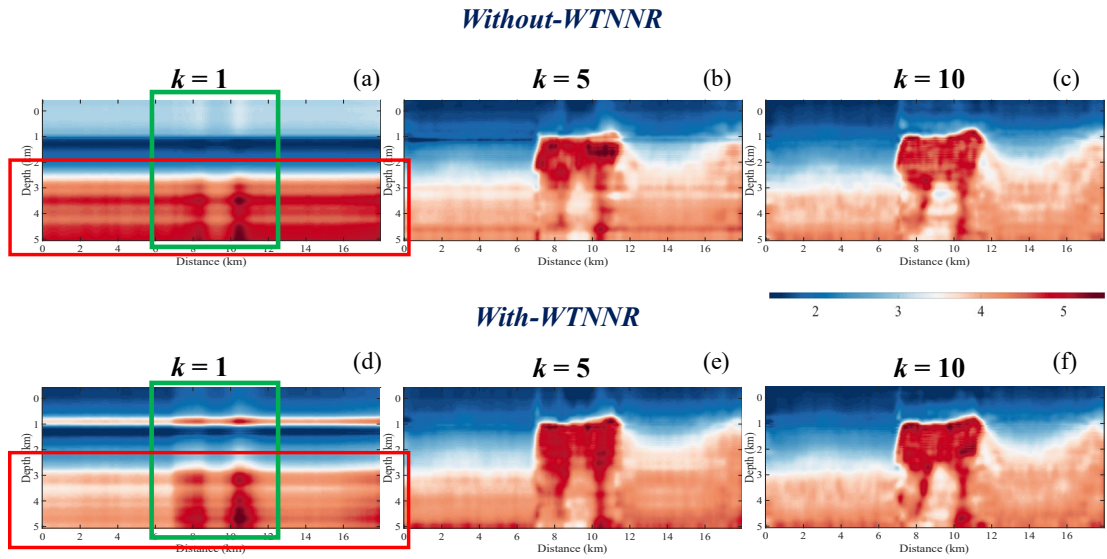

**Figure 7.** (a-c) Test results of velocity increment based on rSVD optimization without-WTNNR with truncation parameters *k=1, k=5,* and *k=10*; (d-f) test results of velocity increment based on rSVD optimization with-WTNNR. WTNNR can effectively enhance the features of the salt columns and slow down the disturbance of the channels on both sides.

**Algorithm 1.** FWI based on the rSVD - WTNNR & iALM.

**Input**: Real seismic data and Initial velocity data, model size $m \times n$, minimum frequency $f_{min}$, maximum frequency $f_{max}$, parameter $t$, truncation parameter $k$, truncation step -size $\varsigma$.

**Initialization:** $f = f_{min}$, $k = 1$, $t_1 = 1$,

    **for** inversion frequency: $f = f_{min} : f_{max}$

    Accelerated iterative optimization,

    where $\lambda_j = \hat{\lambda}_j - \tau(A(\chi)u - b)$,

    $t_{j+1} = (1 + \sqrt{1 + 4(t_j)^2}) / 2$,

    $\hat{\lambda}_{j+1} = \lambda_j + \dfrac{t_j - 1}{t_j + 1}(\lambda_j - \lambda_{j-1}) + \dfrac{t_j}{t_{j+1}}(\lambda_j - \hat{\lambda}_j)$,

        **for** not converged **do**

        $k = k + \varsigma$,

        Compute rSVD: $\left[ \psi_*, \Sigma_*, V_*^* \right] \leftarrow \text{rSVD}(\chi^{-m \times n}, T)$,

        $\chi^{'m \times k} \leftarrow \chi^{-m \times m} \Omega^{m \times k}$,

        where $\Omega$ is Gaussian random matrix,

        Economic QR Decomposition: $Q^{m \times k} \leftarrow eqr(\chi^{'m \times k})$,

        $\chi_c^{k \times n} \leftarrow Q^{*k \times m} \chi^{-m \times n}$,

        $\psi_k^{-k \times k} \Sigma_k^{k \times k} V_k^{*k \times n} \leftarrow \text{SVD}(\chi_c^{k \times n})$,

        where $\tilde{\Psi} = (\psi_1, \ldots \ldots \psi_k) \in \mathbb{R}^{k \times k}$, $V^* = (V_1, \ldots \ldots V_k) \in \mathbb{R}^{k \times n}$,

        Recover left singular vectors: $\psi^{-m \times k} \leftarrow \text{recover}(\psi^{-k \times k})$,

        weighted truncated nuclear norm regularization: $\Sigma_k^{k \times k} \leftarrow \Sigma_k^{k \times k} - \Sigma_{TNNR}^{k \times k}$,

        $\chi^{-m \times n} = \psi^{m \times k} \tilde{\Sigma}_k^{k \times k} V^{*k \times n}$,

        **end for**

    Model update,

    Switch frequency,

    **Reset** $k = 1$,

    **end for**

**Output:** Inversion result.

**Algorithm 1.** FWI based on the rSVD-WTNNR & iALM.

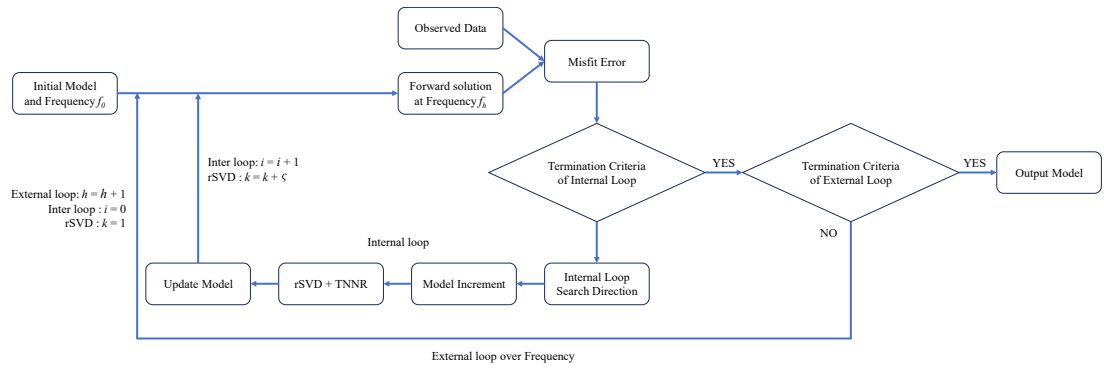

**Figure 8.** Flowchart of accelerated augmented Lagrangian full-waveform inversion based on truncated randomized singular value decomposition in the frequency domain.

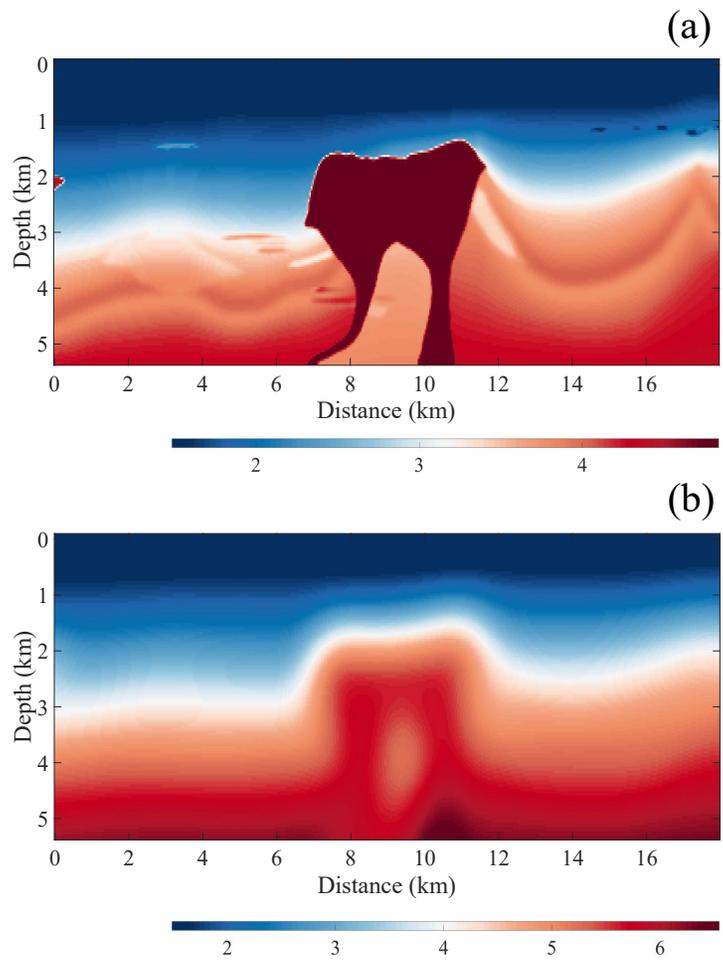

**Figure 9.** 2004 BP model, (a) true velocity model; (b) initial velocity model.

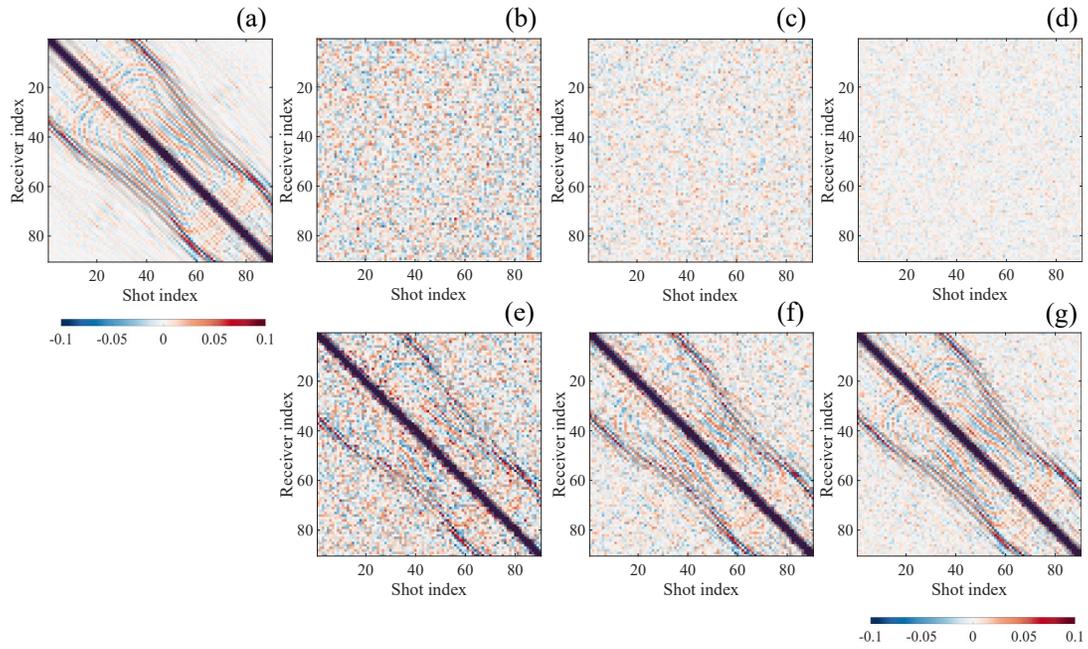

**Figure 10.** The real part of the 3 Hz data for the 2004 BP model. (a) noise-free data; (b-d) random noise of 8 *dB*, 12 *dB*, and 16 *dB*; (e-g) noisy data.

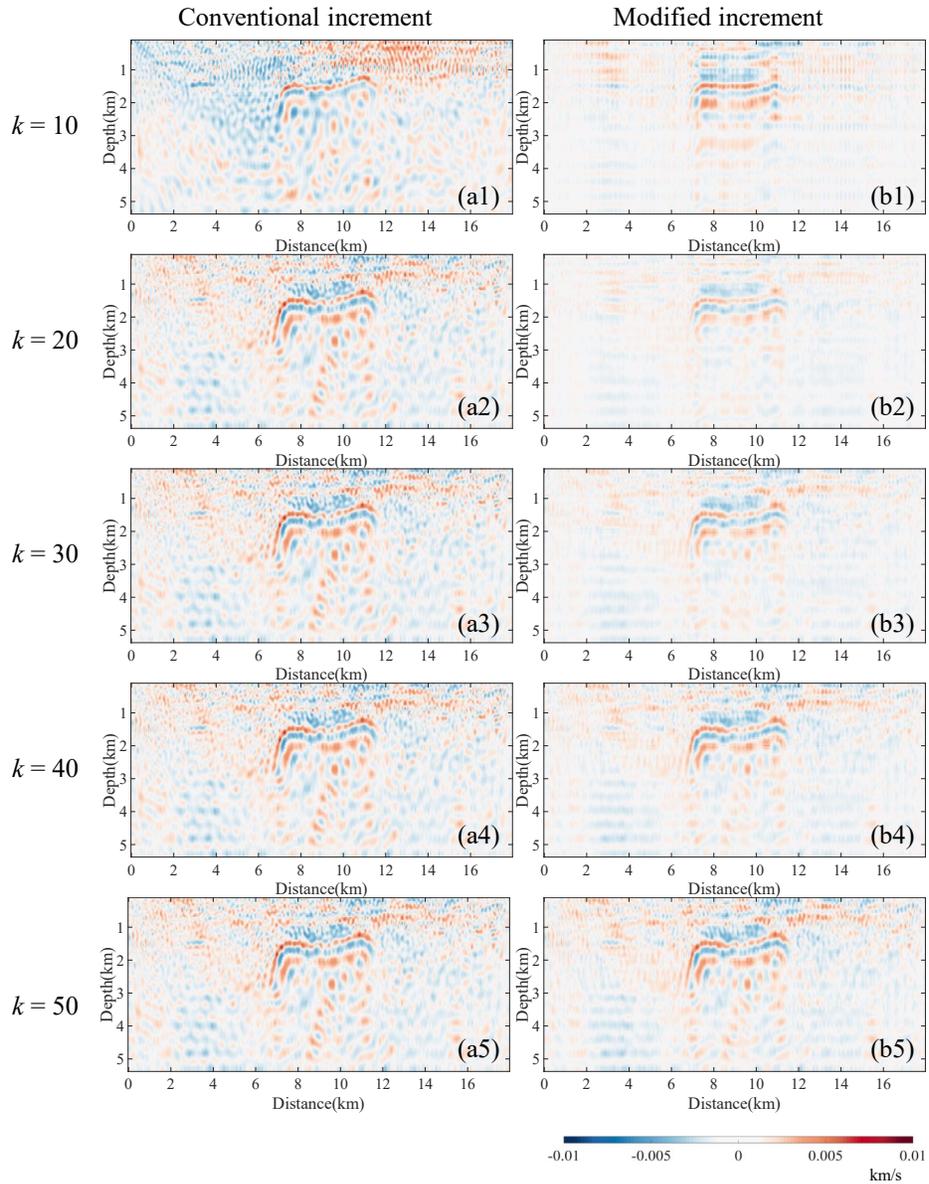

**Figure 11.** Conventional velocity increment model, (a1-a5) close to the 4.5 Hz frequency, the truncation parameter $k$, which increases as the number of internal iterations increases, is equal to 10, 20, 30, 40, and 50, respectively; modified velocity increment model, (b1-b5) close to the 4.5 Hz frequency, the truncation parameter $k$ is equal to 10, 20, 30, 40, and 50, respectively.

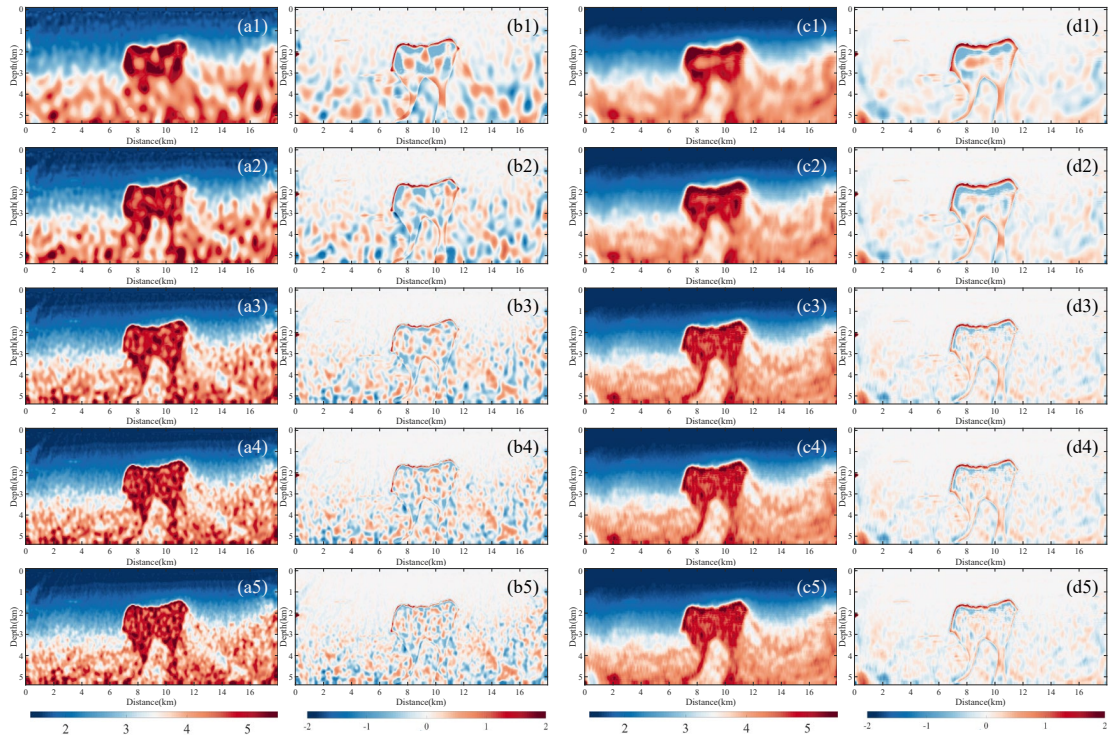

**Figure 12.** 2004 BP model with 8 *dB* random background noise, (a1-a5) inversion results based on Tikhonov regularised FWI with frequencies of 2.15 Hz, 3.9 Hz, 5.35 Hz, 7.70 Hz, and 9.24 Hz, respectively; (b1-b5) velocity differences between the inversion results based on Tikhonov regularised FWI and the true velocity model; (c1-c5) inversion results based on modified FWI with frequencies of 2.15 Hz, 3.9 Hz, 5.35 Hz, 7.70 Hz, and 9.24 Hz, respectively; (d1-d5) velocity differences between the inversion results based on modified FWI and the true velocity model.

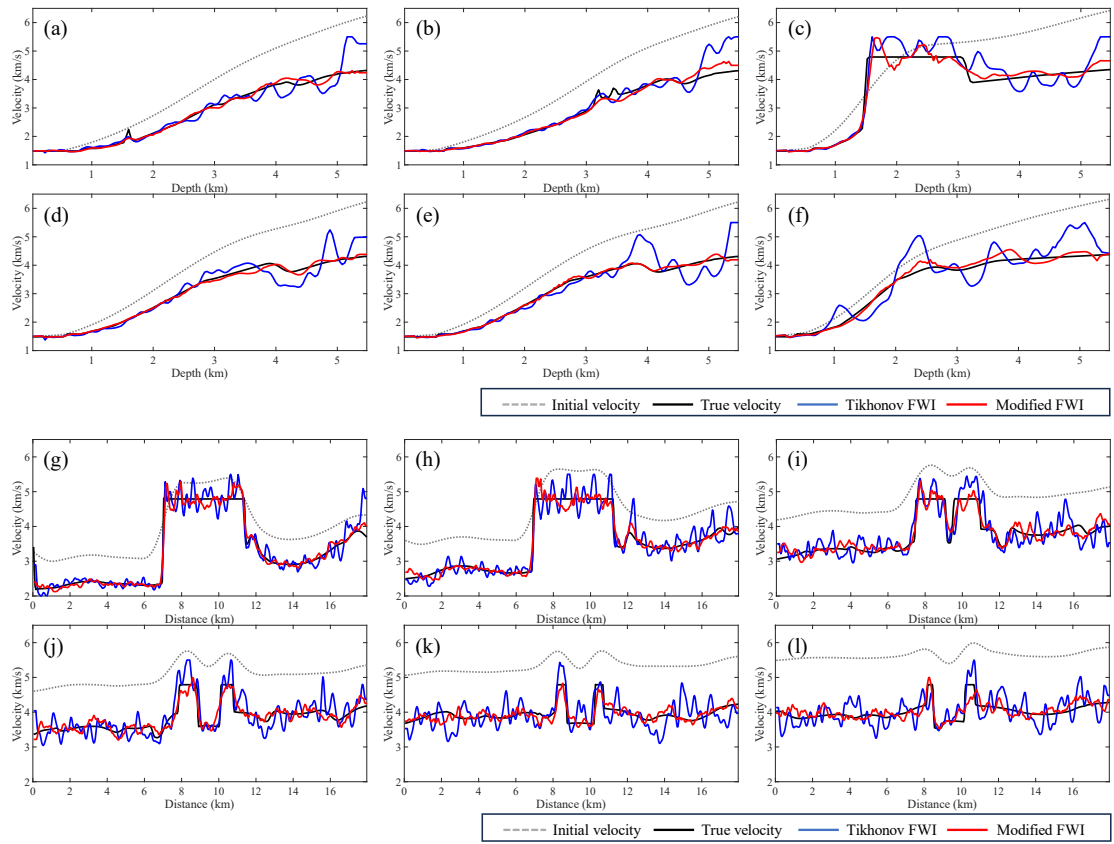

**Figure 13.** 2004 BP model with 8 *dB* random background noise, (a-f) the vertical comparison of one-dimensional velocity models at different *x*-positions, (a) *x* = 2.84 km, (b) *x* = 5.75 km, (c) *x* = 11.08 km, (d) *x* = 13.92 km, (e) *x* = 15.00 km, (f) *x* = 17.80 km; (g-l) the horizontal comparison of one-dimensional velocity models at different *y*-positions, (g) *y* = 2.36 km, (h) *y* = 2.76 km, (i) *y* = 3.36 km, (j) *y* = 3.68 km, (k) *y* = 4.08 km, (l) *y* = 4.56 km; where the actual velocity model is the solid black line, initial velocity model is grey dotted line, the Tikhonov FWI is solid blue line, and the modified FWI is solid red line.

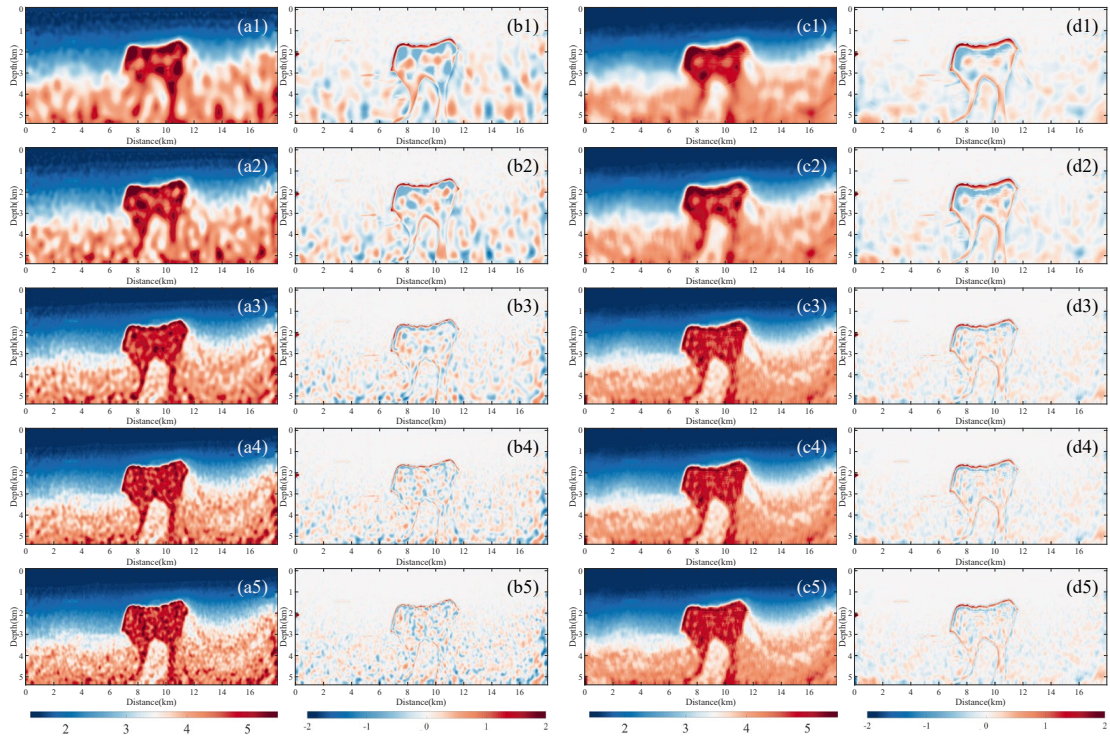

**Figure 14.** 2004 BP model with 12 *dB* random background noise, (a1-a5) inversion results based on Tikhonov regularised FWI with frequencies of 2.15 Hz, 3.9 Hz, 5.35 Hz, 7.70 Hz, and 9.24 Hz, respectively; (b1-b5) velocity differences between the inversion results based on Tikhonov regularised FWI and the true velocity model; (c1-c5) inversion results based on modified FWI with frequencies of 2.15 Hz, 3.9 Hz, 5.35 Hz, 7.70 Hz, and 9.24 Hz, respectively; (d1-d5) velocity differences between the inversion results based on modified FWI and the true velocity model.

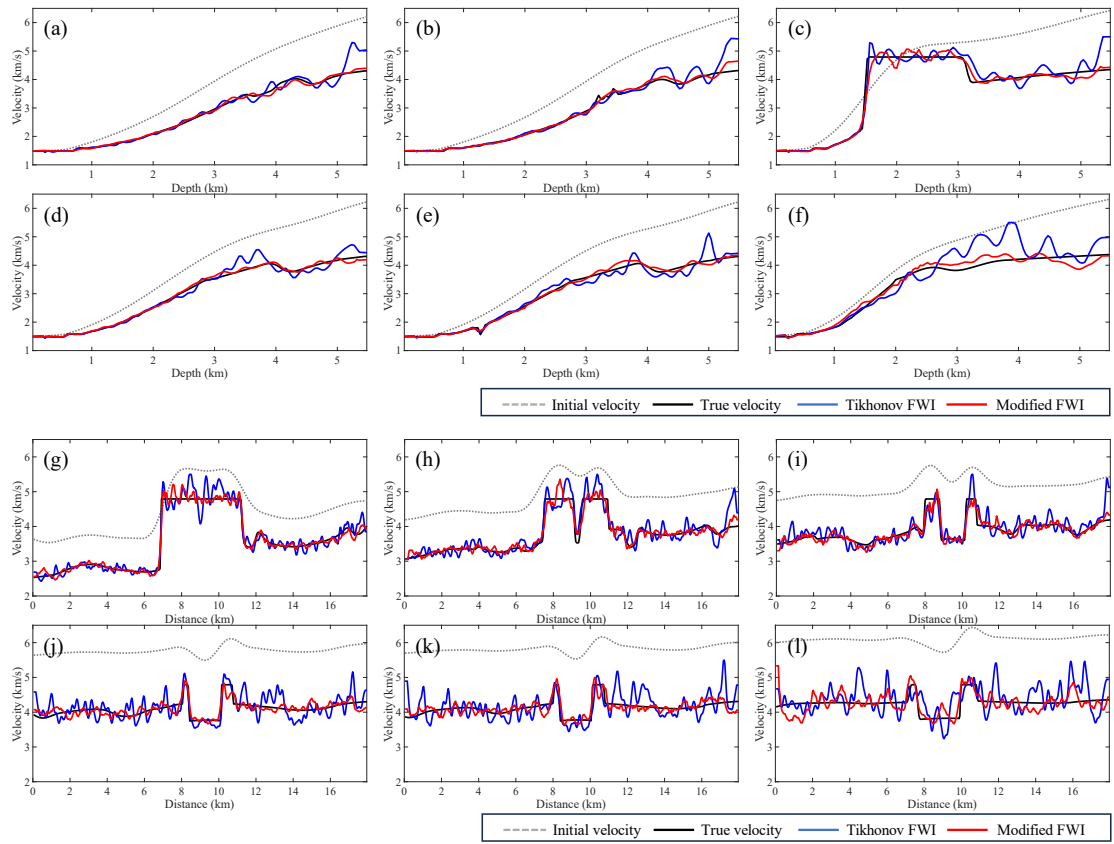

**Figure 15.** 2004 BP model with 12 *dB* random background noise, (a-f) the vertical comparison of one-dimensional velocity models at different *x*-positions, (a) *x* = 4.48 km, (b) *x* = 6.00 km, (c) *x* = 11.08 km, (d) *x* = 13.88 km, (e) *x* = 14.68 km, (f) *x* = 17.76 km; (g-l) the horizontal comparison of one-dimensional velocity models at different *y*-positions, (g) *y* = 2.80 km, (h) *y* = 3.36 km, (i) *y* = 3.80 km, (j) *y* = 4.76 km, (k) *y* = 4.84 km, (l) *y* = 5.28 km; where the actual velocity model is the solid black line, initial velocity model is grey dotted line, the Tikhonov FWI is solid blue line, and the modified FWI is solid red line.

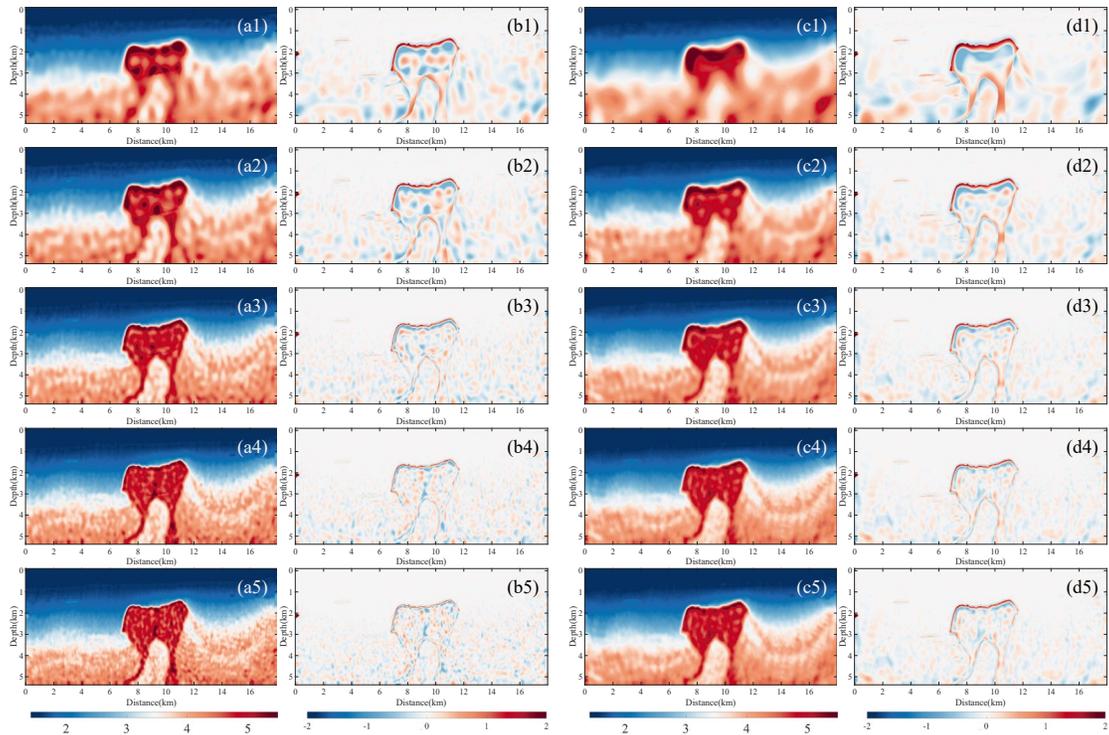

**Figure 16.** 2004 BP model with 16 *dB* random background noise, (a1-a5) inversion results based on Tikhonov regularised FWI with frequencies of 2.15 Hz, 3.9 Hz, 5.35 Hz, 7.70 Hz, and 9.24 Hz, respectively; (b1-b5) velocity differences between the inversion results based on Tikhonov regularised FWI and the true velocity model; (c1-c5) inversion results based on modified FWI with frequencies of 2.15 Hz, 3.9 Hz, 5.35 Hz, 7.70 Hz, and 9.24 Hz, respectively; (d1-d5) velocity differences between the inversion results based on modified FWI and the true velocity model.

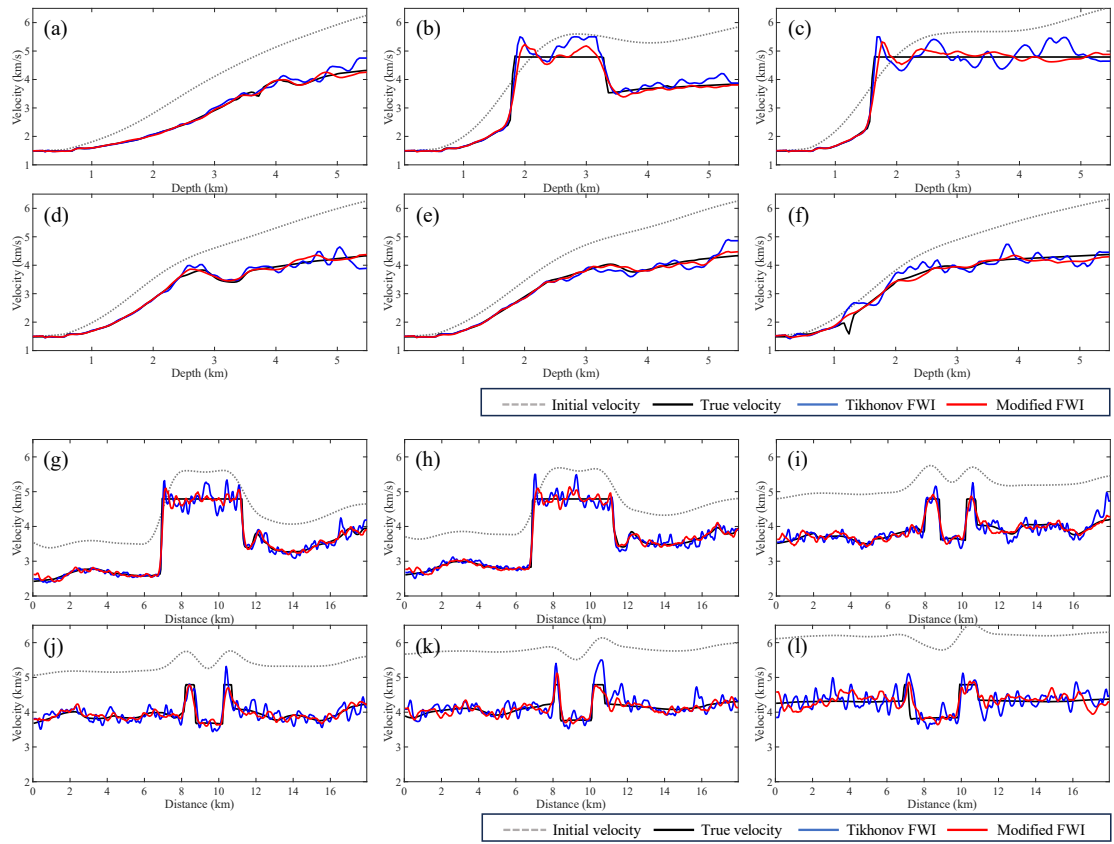

**Figure 17.** 2004 BP model with 16 *dB* random background noise, (a-f) the vertical comparison of one-dimensional velocity models at different *x*-positions, (a) *x* = 6.44 km, (b) *x* = 9.24 km, (c) *x* = 10.44 km, (d) *x* = 12.12 km, (e) *x* = 16.20 km, (f) *x* = 18.00 km; (g-l) the horizontal comparison of one-dimensional velocity models at different *y*-positions, (g) *y* = 2.68 km, (h) *y* = 2.88 km, (i) *y* = 3.84 km, (j) *y* = 4.08 km, (k) *y* = 4.80 km, (l) *y* = 5.44 km; where the actual velocity model is the solid black line, initial velocity model is grey dotted line, the Tikhonov FWI is solid blue line, and the modified FWI is solid red line.

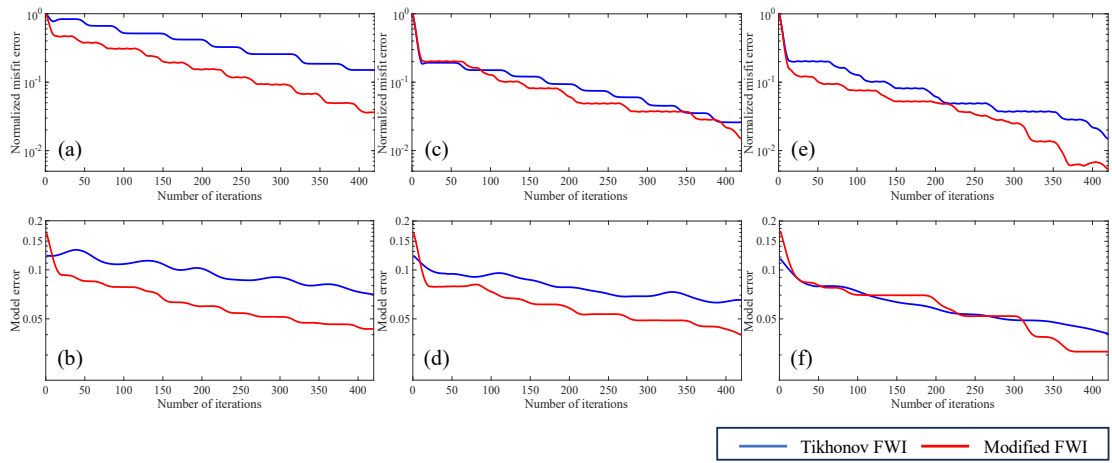

**Figure 18.** Model curves for the test, (a-b) misfit error and model error in the case of 8 *dB* random background noise; (c-d) misfit error and model error in the case of 12 *dB* random background noise; (e-f) misfit error and model error in the case of 16 *dB* random background noise; the solid red line represents the modified FWI, and the blue line represents the Tikhonov FWI.

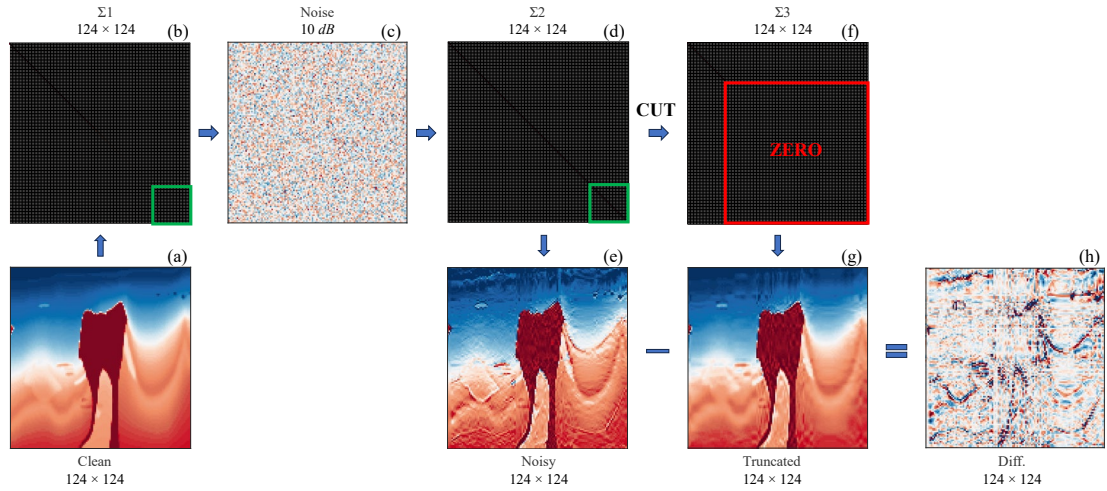

**Figure 19.** (a) The initial test velocity increment; (b) the singular value matrix for this velocity increment; (c) 10 *dB* of random noise; (d) the test velocity increment after adding the noise disturbance; (e) the velocity increment after reconstructing the singular value matrix (d); (f) the new singular value matrix after simply truncating the singular value matrix (d); (g) the new velocity increment with a reconstructing of the singular value matrix (f); and (h) the difference between velocity increments (e) and (g).

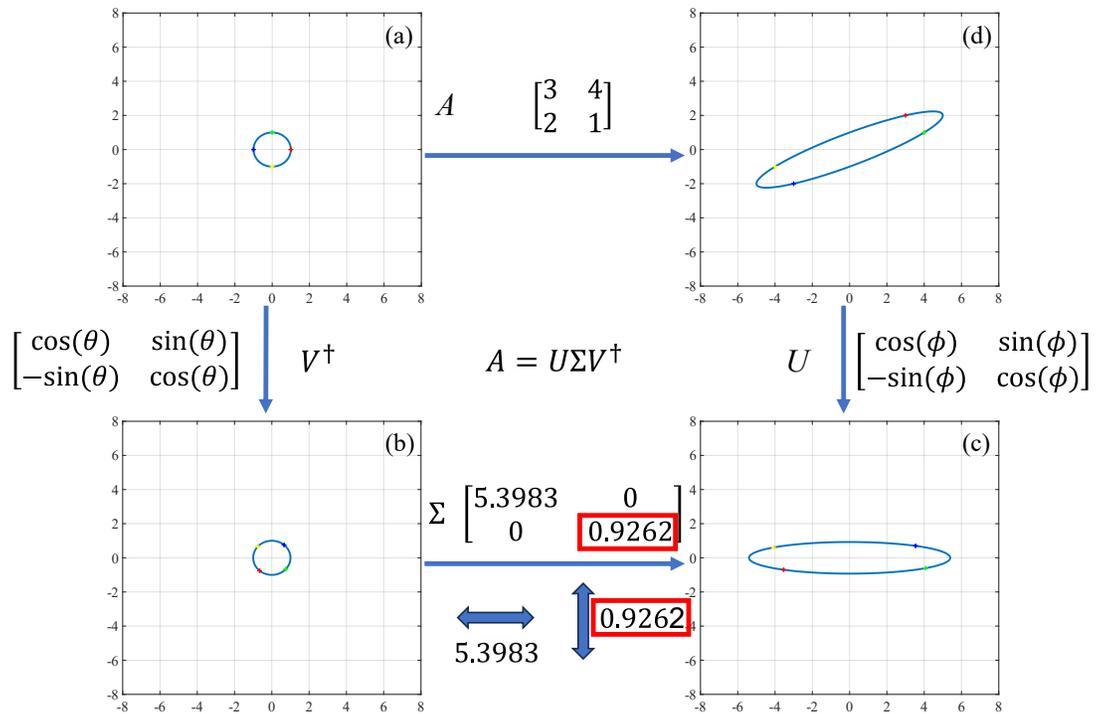

**Figure 20.** Geometric interpretation of SVD, (a) target matrix; (b) left orthogonal matrix representing rotation; (c) singular value matrix representing longitudinal and horizontal stretching; (d) right orthogonal matrix representing rotation.

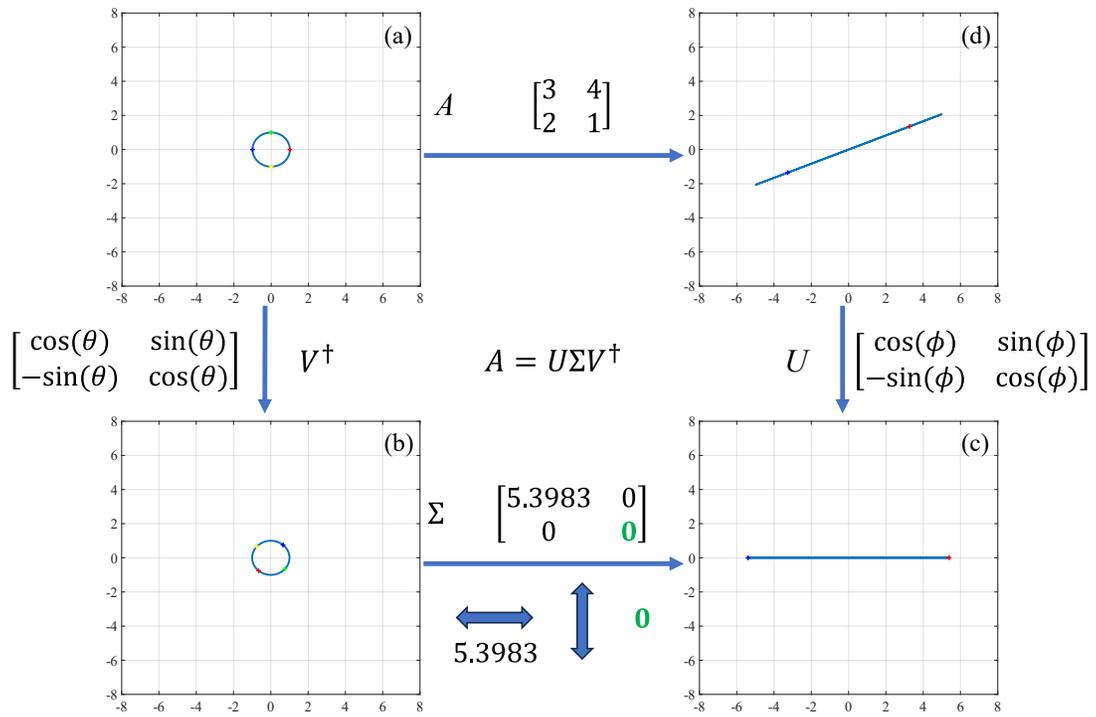

**Figure 21.** Geometric interpretation of the truncation operation of the SVD, (a) the target matrix; (b) the left orthogonal matrix represents the rotation; (c) the singular value matrix represents the longitudinal and horizontal stretching and truncation of its smallest singular value results in a longitudinal stretch of 0; (d) the right orthogonal matrix represents the rotation.